\documentclass[fleqn,usenatbib]{mnras}
\usepackage{amsmath}
\usepackage{graphicx}
\usepackage{booktabs,caption}
\usepackage[flushleft]{threeparttable}
\usepackage{xstring}
\usepackage{xcolor}

\usepackage[T1]{fontenc}
\usepackage{ae,aecompl}

\usepackage{newtxtext,newtxmath}

\newcommand{\ium}{$\micron^{-1}$}
\newcommand{\nodata}{.\,.\,.}
\newcommand{\cbeta}{$c(\mathrm{H\beta})$}

\newcommand{\ebv}{$E(B-V)$}

\title[FUV study of C$_{60}$-PNe]{An ultraviolet spectral study of fullerene-rich planetary nebulae}

\author[G{\'o}mez-Mu{\~n}oz et al.]{
M.~A. G{\'o}mez-Mu{\~n}oz,$^{1,2}$
\thanks{Contact e-mail: \href{mailto:magm@iac.es}{magm@iac.es}}
D.~A. Garc{\'\i}a-Hern{\'a}ndez,$^{1,2}$
A. Manchado,$^{1,2,3}$
R. Barzaga,$^{1,2}$ \vspace{4pt}\\
{\Large \normalfont and T. Huertas-Rold{\'a}n$^{1,2}$}
\\
$^1$Instituto de Astrof{\'i}sica de Canarias (IAC), E-38205 La Laguna, Tenerife, Spain\\
$^2$Departamento de Astrof{\'i}sica, Universidad de La Laguna (ULL), E-38206 La Laguna, Tenerife, Spain\\
$^3$Consejo Superior de Investigaciones Cient{\'i}ficas (CSIC), Spain
}

\date{\today}
\pubyear{2023}

\begin{document}
\label{firstpage}
\pagerange{\pageref{firstpage}--\pageref{lastpage}}
\maketitle

\begin{abstract}
Several planetary nebulae (PNe) have been found to contain both polycyclic aromatic hydrocarbon (PAH-like) species and fullerenes (C$_{60}$) distinguished by their mid-infrared emission. Previous laboratory and astronomical studies suggest that the formation of both species could be related to the decomposition, by photochemical processing, of hydrogenated amorphous carbon (HAC) grains. Then, HACs and, seemingly, big-fullerene related species (e.g., carbon onions) have been suggested as potential carriers of the UV bump at 2175 \AA~and the far-UV rise common to interstellar extinction curves. Our goal is to investigate the UV bump with the possible presence of a HAC extinction component in the International Ultraviolet Explorer (IUE) spectra of C-rich PNe; both with detected and non-detected fullerenes. The considered sample includes three C$_{60}$-PNe (Tc 1, IC 418, and IC 2501) and the non-C$_{60}$-PN Hen 2-5. Independently of the presence of C$_{60}$ in their circumstellar envelopes, we found that the UV bump in all sample PNe is well explained by interstellar extinction, suggesting that species different from those of the foreground insterstellar medium, e.g., large fullerene-related species like carbon onions, are not the carrier. Interestingly, we found that PNe Tc 1 and Hen 2-5 show an absorption in the FUV rise. Their IUE continuum spectra may be very well reproduced by including the extinction curve of HAC-like very small grains (VSG). The possible presence of both species, HAC-like grains and fullerenes (C$_{60}$), in Tc 1 could support the HAC photochemical processing scenario for the formation of fullerenes in the complex circumstellar envelopes of PNe.
\end{abstract}

\begin{keywords}
planetary nebulae: general --- ISM: dust, extinction --- ultraviolet: ISM --- astrochemistry --- circumstellar matter
\end{keywords}



\section{Introduction}

Planetary nebulae (PNe) are one of the latest stages in the evolution of low- and intermediate-mass
stars ($\sim$0.8$-$8.0~M$_{\sun}$), which consist of a ionised envelope surrounding a white dwarf (WD) star.
Most of the dust and gas seen in PNe is formed during the previous asymptotic giant branch (AGB) phase by the removal of the external H shell through a strong mass-loss ( even at $\sim$10$^{-5}$ M$_{\sun}$/yr).
Also, the most important nucleosynthesis processes \citep[the formation of both light and heavy neutron-rich elements;][]{Karakas2014}
take place during the AGB phase, and they completely mark the following post-AGB and PN evolutionary
phases (e.g., C-rich vs O-rich objects).
In fact, it is also believed that most of the dust observed in the local
Universe comes from the mass lost by AGB stars to the insterstellar medium \citep{Ferrarotti2006}. Thus, the study of the dust grains and molecular species in this type of objects turns out to be very important in order to understand the enrichment and chemical composition of the interstellar and intergalactic media.

Carbon-based molecular and solid-state species, which are mainly produced in the envelopes of low-mass AGB stars, are very abundant in the interstellar (ISM) and circumstellar (CSM) matter \citep[][and references therein]{Draine2003,Molster2003,Gail2009,Jager2011}.
In particular, the spectral energy distribution of young C-rich PNe shows the presence of diffuse interstellar bands \citep[DIBs; see e.g.,][for a review]{omont2016} in the optical and near-IR \citep[e.g.][]{Luna2008,GarciaHernandez2013,diazluis2015,Zasowski2015} as well as polycyclic aromatic hydrocarbon (PAH-like) species and fullerenes (such as C$_{60}$ and C$_{70}$) in the mid-IR \citep[e.g.][]{GarciaHernandez2010,GarciaHernandez2011}. Indeed, the most common fullerenes C$_{60}$ and C$_{70}$ were first detected in the \textit{Spitzer} spectra of the young PN Tc~1 \citep{Cami2010}. Nowadays, the presence of C$_{60}$ has been confirmed in many astrophysical objects such as reflection nebulae \citep{sellgren2010}, PNe \citep{GarciaHernandez2010,GarciaHernandez2011, GarciaHernandez2012,Otsuka2014}, peculiar R Coronae Borealis stars \citep[RCB;][]{GarciaHernandez2011b}, post-AGB stars \citep{Zhang2011,Gielen2011}, and Herbig Ae/Be stars \citep{Arun2023}, among others, by its four strongest mid-IR emission features at $\sim$7.0, 8.5, 17.4, and 18.9{\micron}.

The formation route of fullerenes in H-rich environments like PNe is still uncertain. However, two different top-down chemical models towards the most stable C$_{60}$ and C$_{70}$ molecules seem to be the most suitable ones to explain the presence of fullerenes in the circumstellar (CSM) and interstellar media (ISM): i) the photochemical processing of hydrogenated amorphous carbon grains \citep[HACs or a-C:Hs; ][]{GarciaHernandez2010}; and ii) the photochemical processing of large PAHs \citep{Berne2012,Murga2022}. Interestingly, \citet{Duley2012} suggested that
fullerenes are only observed in objects with extreme radiation such as PNe \citep{Otsuka2014} and Herbig Ae/Be stars \citep[where C$_{60}$ is only detected in the B-type stars;][]{Arun2023}, whereas in cooler objects like RCB and post-AGB stars we are only seeing proto-fullerenes \citep{Duley2012}; i.e. precursor molecules that subsequent react to yield C$_{60}$\footnote{Proto-fullerenes are molecules similar to C$_{60}$ but with different shapes (no icosahedral structure) that can be transformed into C$_{60}$ during the de-hydrogenation of HACs (a-C:H $\rightarrow$ a:C + proto-fullerenes $\rightarrow$ C$_{60}$ + other fullerenes) due to a strong radiation; a suggested evolutionary sequence, based on laboratory experiments, to convert hydrogen-rich carbons into fullerenes (see e.g., \citealt{Duley2012} and \citealt{Jager2011}).}.

The ISM and CSM extinction curves contain crucial information about the composition and size distribution of the dust grains. The construction of dust models to explain their extinction curves is a useful tool to study the dust chemical composition. These extinction curves are derived from observations of several lines of sight from the near-IR to the UV wavelength range. A strong absorption bump at 4.6{\ium} ($\sim$2175{\AA}) is a persistent spectral feature \citep{Stecher1965,Wickramasinghe1965}.
Its central position is constant over several lines of sight and its origin is still largely debated \citep[][and references therein]{Fitzpatrick1999,Fitzpatrick2007}. Also, many of these extinction curves have a steep far-ultraviolet (FUV) rise, tracing extinction variations.
The absence of a relationship between the UV bump's spectral properties and the relative strengths of the bump
and the FUV rise indicates that there may be different types of dust carriers
\citep{Greenberg1983,Fitzpatrick2007,CecchiPestellini2008,Gavilan2016,Gavilan2017}.
Graphite particles were initially proposed as the carrier
of the UV bump \citep{Wickramasinghe1965,Draine1984}. However, in the same way, complex carbon-based molecular species have also been proposed as potential UV bump carriers; e.g., carbon onions \citep{IglesiasGroth2004} and hydrogenated fullerenes \citep{Cataldo2009}.
More recently, \citet[][and references therein]{Massa2022} suggest that PAHs can also contribute to the UV Bump due to the strong correlation found between the strength of the bump and the PAH emission bands. Subsequently, \citet{Lin2023} investigated the contribution of PAHs to the UV bump by using quantum chemical calculations, resulting in a slightly different position and width to what is observed.
On the other hand, the contribution of more disordered species to the variations of the UV Bump and FUV rise was investigated by introducing the HACs, soot particles \citep[][and references therein]{Gadallah2011,Gavilan2016,Gavilan2017}, and other carbonaceous materials such as flakes, pretzels, and cages \citep{Dubosq2019,Dubosq2020}.

In particular, \citet{Gavilan2016,Gavilan2017} argued that soot particles are suitable
carriers of the UV bump, while more hydrogenated carbon materials (like HACs) contribute to the 
steep FUV rise. In the same way, \citet{Dubosq2020} showed that cages and flakes are likely 
contributors to the soot constituents and possible carriers of the interstellar UV bump.

In this study we aim to explore the UV bump and the FUV rise in C-rich PNe; both
with detected and non-detected fullerenes. We investigate the existence of carbon analogues in the circumstellar envelopes of four C-rich PNe
by analyzing their ISM and CSM extinction curves using their UV spectra.
The paper is structured as follows: section~\ref{sec:observations} introduces the sample and the stellar parameters of PNe used in our analysis.
Section~\ref{sec:analysis} presents an analysis of their UV spectra using an average Milky-Way (MW) extinction curve.
Section~\ref{sec:discusion} focuses on the analysis and discussion of including carbon analogues in the extinction curves.
Finally, a summary is provided in Section~\ref{sec:summary}.

\section{Observational data}
\label{sec:observations}

In order to study the UV spectra of fullerene-rich PNe,
we downloaded the best \textit{International Ultraviolet Explorer} (IUE)
spectra from the \textit{IUE newly extracted spectra}
(INES\footnote{\url{http://sdc.cab.inta-csic.es/ines/}}) archive data.
We looked for all fullerene-containing (C$_{60}$) PNe, from \citet{GarciaHernandez2012}
and \citet{Otsuka2014},
for which both the short-wavelength
(SW 1150--2000{\AA}) and long-wavelength (LW 1850--3300{\AA}) spectra
were available; we downloaded only those taken in low-resolution mode ($\sim6${\AA}) and
with the large aperture (10$\times$20{\arcsec}) to include most of the PN emission\footnote{Unfortunately, there are no high-quality IUE spectra with smaller slits for the sample of C$_{60}$-PNe studied here. Note, however, that the nebular contribution in Tc 1 and Hen 2-5 is marginal and the continuum seen in the IUE spectra can be thus assumed to be completely dominated by the central star (see Figure~\ref{fig:pne_nebcont}).}. For comparison, we also looked for C-rich dust PNe (with non-detected fullerenes) from
\citet{Stanghellini2012}. A total of three (IC~418, IC~2501, and Tc~1) and one (Hen~2-5) PNe,
with detected and non-detected fullerenes, respectively, were found in the INES archive.

The sample mainly consist of moderate reddened (E$(B-V)$=0.11--0.4) PNe with
similar effective temperatures, $T_\mathrm{eff}$, and sizes$<$15{\arcsec}.
The sample of PNe together with their properties ($T_\text{eff}$, luminosity, gravity, and diameter) are shown in Table~\ref{tab:pne_iue}.

\begin{table*}
    \centering
    \begin{threeparttable}
    \caption{Properties of the PNe and CSPNe obtained from the literature and the IUE spectra obtained from the INES database. \label{tab:pne_iue}}
    \begin{tabular}{llccrrrrcc}
    \toprule
    PN~G & Name & RA & DEC & $T_\text{eff}$ & $\log (L/L_{\sun})$ & $\log g$ & Diameter$^{a}$ & IUE & Ref. \\
        &      &    &     & (kK)       &    & (cm\,s$^{-2}$)   & (arcsec) &  &  \\
    \midrule
    215.2$-$24.2 & IC~418 & 05:27:28.21 & $-$12:41:50.28 & 37 & 3.88 & 3.55 & 14.0 & SWP03178 LWR03390 & 1 \\
    264.4$-$12.7 & Hen~2-5 & 07:47:20.04 & $-$51:15:03.43 & 40 & 3.54 & \nodata & 3.8 & SWP29584 LWP09453 & 2 \\
    281.0$-$05.6 & IC~2501 & 09:38:47.19 & $-$60:05:30.81 & 55 & 3.47 & 4.75 & 8.0 & SWP16319 LWR12566 & 3,4 \\
    345.2$-$08.8 & Tc~1 & 17:45:35.29 & $-$46:05:23.72 & 30 & 3.58 & 3.40 & 12.9 & SWP13413 LWP17557 & 4,5 \\
    \bottomrule
    \end{tabular}
    \begin{tablenotes}
    \item $^{a}$ Diameters were extracted from the Hong Kong/Australian Astronomical Observatory/Strasbourg Observatory H$\alpha$ PN database \citep[HASH; ][]{Parker2016}.
    \item \textbf{Refs.}: (1) \citet{Morisset2009} -- (2) \citet{MorenoIbanez2016} -- (3) \citet{Ali2019} -- (4) -- \citet{Otsuka2014} -- (5) \citet{Aleman2019}
    \end{tablenotes}
    \end{threeparttable}
\end{table*}

\section{Results}
\label{sec:analysis}

\begin{figure}
    \centering
    \includegraphics[width=\columnwidth]{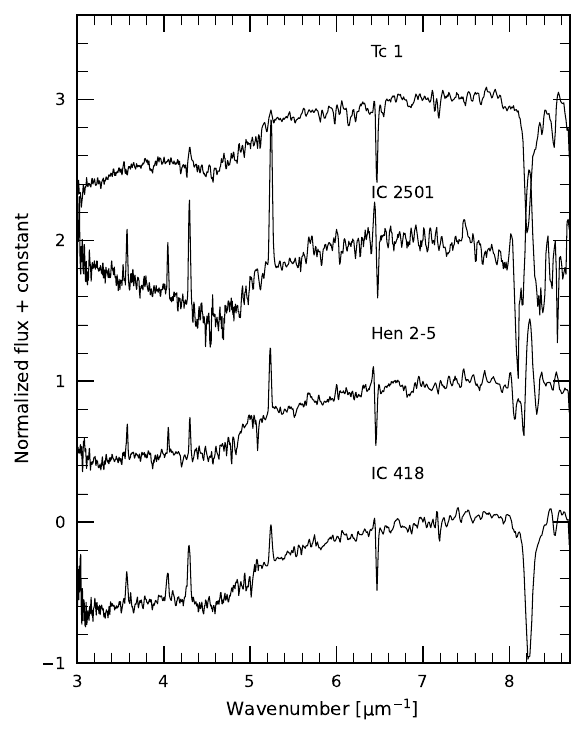}
    \caption{IUE spectra of our sample of PNe. The spectra are composed of both
    the SW and LW matched at 5.13{\ium}. A three-point smooth was applied on each spectrum. \label{fig:all_sample}}
\end{figure}

Figure~\ref{fig:all_sample} shows the IUE spectra of the four C-rich PNe in our sample (detailed information
is presented in Table~\ref{tab:pne_iue}). The SW and LW IUE spectra were matched at 1950{\AA} to reconstruct the full UV spectrum for each PN; a three-point smooth was also applied to each reconstructed spectrum. A strong UV bump (at $\sim$2175 {\AA} or 4.6 {\ium}) can be seen in all sample PNe, being particularly strong in IC\,2501. 

Traditionally, the extinction of a PNe is obtained by measuring the observed
H Balmer emission lines and comparing its value with theoretical values
(specially comparing the observed and theoretical values of H$\alpha$:H$\beta$:H$\gamma$);
the expected theoretically ratio of a series of Balmer lines can be found with a single
value of E($B-V$) \citep[see][for a review]{Cahn1992,Osterbrock2006}. Another method
is by comparing the radio continuum flux density with the H$\beta$ flux in which
the flux densities have the same dependence in electronic density, $n_\mathrm{e}$, and
then the expected ratio is a weak function of electronic temperature, $T_\mathrm{e}$,
and He abundance \citep[e.g.][]{Cahn1992}.
The extinction can also be calculated by measuring the sharp peak near 2175{\AA}
(4.6{\ium}; see Fig.~\ref{fig:all_sample}; the so-called UV bump), with a half-width of $\sim$400{\AA}
($\sim$0.84{\ium}), by varying {\ebv}
to remove the absorption peak.
As our intention is to analyse the UV extinction (both the UV bump and the FUV rise components)
by including the stellar and nebular continuum emission,
we need to obtain the {\ebv} value with an indirect method independent of the UV bump. Therefore, the first method (i.e. the comparison observations vs. theory for the H Balmer lines) was used to analyse the extinction towards the four sample C-rich PNe presented in Table~\ref{tab:pne_iue}. For each PN,  the observed nebular emission lines, physical parameters and ionic abundances from literature are listed in Table~\ref{tab:physics_abundances}.

As mentioned in Sec.~\ref{sec:observations}, the selected IUE spectra were taken with the
larger aperture (10$\times$20{\arcsec}) in order to include most of the PN emission.
Thus, in the analysis of the UV spectra,
we have to take into account that the observed spectrum is a composition of the
stellar (SA) and nebular continuum (NC).
We used the theoretical post-AGB stellar
atmosphere models from TMAP archive\footnote{The models were extracted from the theoretical
atmosphere models database from the SVO theoretical services.
\url{http://svo2.cab.inta-csic.es/theory/newov2/}},  a theoretical pure H non-local thermodynamic equilibrium
(NLTE) model stellar atmospheres calculated using the code TMAW \citep{Rauch2018},
to reproduce the stellar continuum using the $T_\mathrm{eff}$ and $\log g$ reported by different authors
for each PN (see Table~\ref{tab:pne_iue}).
We used the NEBCONT code under the {\sc dipso}
package of the Starlink v2021A\footnote{\url{http://starlink.eao.hawaii.edu/starlink}} \citep{Currie2014,Berry2022}
to generate the NC emission by adopting the $\log F(\mathrm{H}\beta)$, 
$T_\text{e}$, $n_\text{e}$, and ionic abundances reported in the
literature (see Table~\ref{tab:physics_abundances}).
The observed UV spectra were unreddened using the extinction
law of \citet[hereafter FM07]{Fitzpatrick2007}\defcitealias{Fitzpatrick2007}{FM07}, which includes an update
in the parametrization of the UV range
of the original extinction curve of \citet{Fitzpatrick1999},
using the {\cbeta} reported in Table~\ref{tab:physics_abundances}.
The SA, in units of erg\,s$^{-1}$\,cm$^{-2}$\,{\AA}$^{-1}$, was scaled by $\theta$ (where $\theta = (R/D)^{2}$, with $R$ and $D$ being the radius of the central star and the distance to the CSPN, respectively) in such a way that the overall flux matched the continuum around the UV bump
\citep[a similar analysis was carried out for NGC~2346 and SaSt~2-3 in][respectively]{GomezMunoz2019,Otsuka2019}.
We also fixed $R(V)$ to the average Galactic value of 3.1\footnote{Other values of $R(V)$ were used following the \citet{CCM89} extinction law, but none have been successful (see Appendix~\ref{ap:rv_vary}). Thus, we opted to keep it simple by implementing the updated UV averaged parameterised form of the \citetalias{Fitzpatrick2007} extinction curve.}. The following equation is best describing
the model we fitted to the observed spectrum,
\begin{equation}
\label{eq:ebv_vary}
    F_\mathrm{\lambda^{-1}} = SA_\mathrm{\lambda^{-1}}\theta + NC_\mathrm{\lambda^{-1}}
\end{equation}
where $\theta$, is varied in such a way that the $F_{\lambda^{-1}}$ match the corrected UV bump.

\begin{figure*}
    \centering
    \includegraphics[width=\columnwidth]{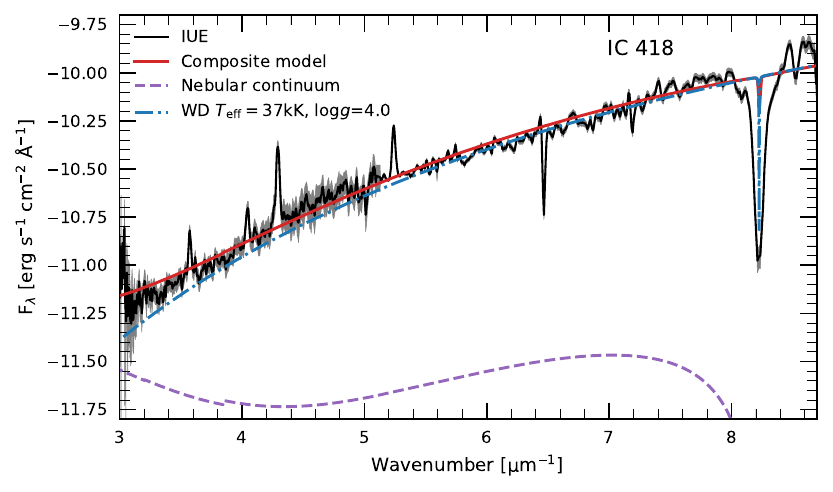}
    \includegraphics[width=\columnwidth]{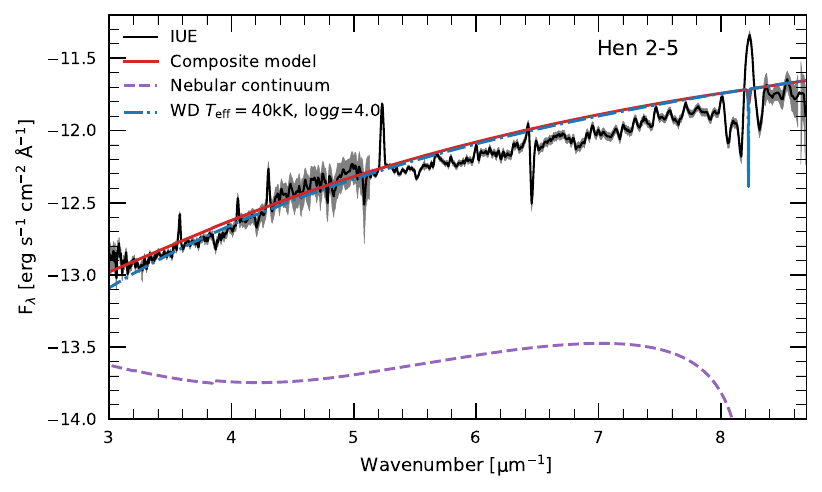}
    \includegraphics[width=\columnwidth]{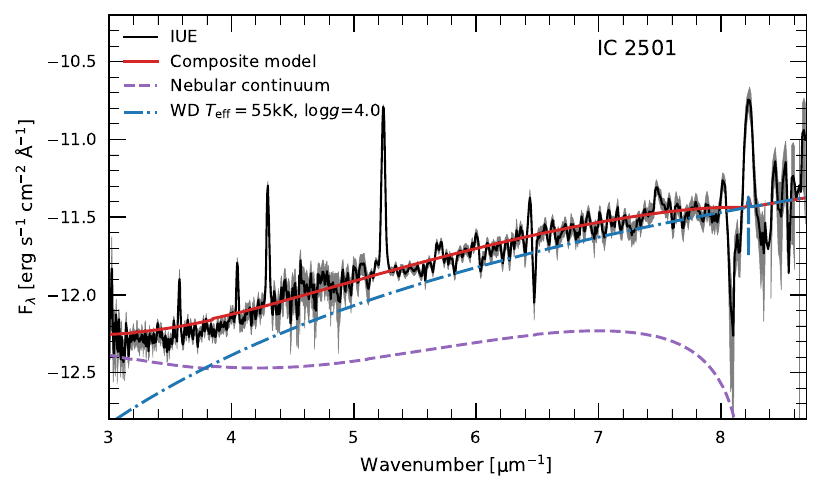}
    \includegraphics[width=\columnwidth]{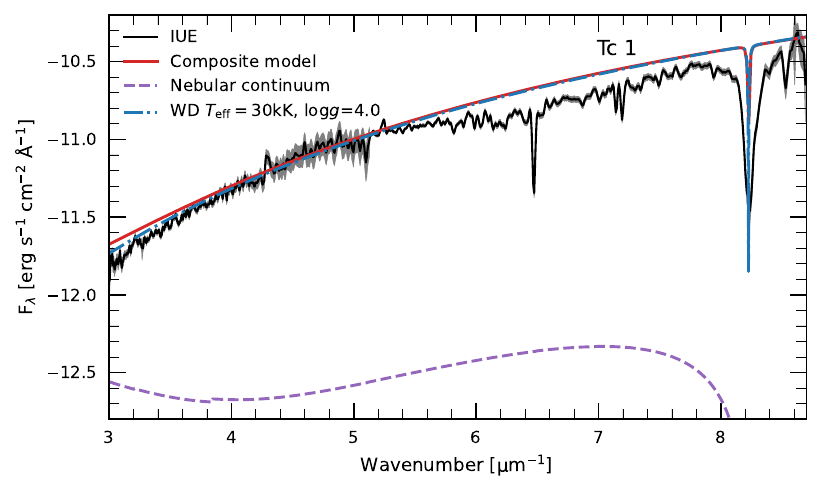}
    \caption{The IUE spectra (black line) of IC~418, Hen~2-5, IC~2501 and Tc~1 (upper-left, upper-right, bottom-left and bottom-right panel, respectively) are shown along with the nebular continuum (purple dashed-line) and stellar atmosphere (blue dash-dotted) models as well as the composite model (red line). The IUE sigma are marked as a filled grey area. \label{fig:pne_nebcont}}
\end{figure*}

Figure~\ref{fig:pne_nebcont} shows all the IUE spectra of the sample PNe 
along with the SA (blue dashed-line), the NC emission (purple dashed-line),
and the composite model (using eq.~\ref{eq:ebv_vary});
the observed IUE spectra were dereddened with the {\cbeta} values
listed in Table~\ref{tab:physics_abundances}.
All PNe appear to be not strongly affected by the
NC emission, with the exception of IC~2501, which displays a strong nebular
continuum emission contribution in the near-UV (NUV) region ($<$5.0{\ium}).
The composite spectrum fits very well the IC~418 and IC~2501 UV observations by
using the physical conditions and ionic abundances from the literature (Table~\ref{tab:physics_abundances}). However, for the PNe Hen~2-5 and Tc~1, the CSPN are extinguished in the far-UV (FUV; $>$5.5{\ium}) with respect to the composite spectrum; although the composite spectrum fits very well in the NUV.
Other values for {\cbeta} have been tried ($\pm$20\% from the literature values; Table~\ref{tab:physics_abundances}), to successful fitting the NUV and FUV regions simultaneously but the fits were not as satisfactory as those obtained with the literature values.
The fact that observations can not be reproduced by the models by applying an averaged extinction curve may indicate a different contribution in the line-of-sight of each PNe; i.e. that the extinction in the UV bump and FUV rise could be due to a different carrier, as suggested by \citet{Greenberg1983}, or/and 
that there is an additional contribution to the extinction from a circumstellar envelope.

It should be noted here that, when using the average extinction curve for the Galaxy, the composite models fit the observed PN UV spectra (independently of the fullerene detections) very well at wavenumbers $<$5.0{\ium}; i.e. the UV bump and the surrounding continuum. This suggests that there is no evidence for other molecular (or solid-state) UV bump carriers \citep[e.g. carbon onions;][]{IglesiasGroth2004}, different from those of the foreground ISM, possibly present in the PNe circumstellar envelopes and affecting or contributing to the UV extinction; at least for the sample PNe presented in this study.

\begin{table}
    \centering
    \resizebox{\columnwidth}{!}{
    \begin{threeparttable}
    \caption{Observed emission lines, physical parameters, and ionic abundances from the literature. \label{tab:physics_abundances}}
    \begin{tabular}{lcccc}
    \toprule
     Name  & IC~418 & Hen~2-5 & IC~2501  & Tc~1 \\
    \midrule
    \multicolumn{5}{c}{Log ionic abundances} \\
    \hline
     He$^{+}$/H$^{+}$ & $-$1.022 & -1.009 & $-$0.987 & $-$0.980 \\
     He$^{+2}$/H$^{+}$ & \nodata & \nodata & $-$4.740 & \nodata \\
     N$^{+}$/H$^{+}$ & $-$4.387 & \nodata & $-$4.940 & $-$4.540 \\
     N$^{+2}$/H$^{+}$ & \nodata & \nodata & \nodata & $-$3.400$^{b}$ \\
     C$^{+}$/H$^{+}$ & \nodata & \nodata & \nodata & \nodata \\
     C$^{+2}$/H$^{+}$ & \nodata & \nodata & $-$3.028$^{\text{b}}$ & $-$3.220$^{b}$ \\
     C$^{+3}$/H$^{+}$ & \nodata & \nodata & $-$4.646$^{\text{b}}$ & \nodata \\
     O$^{+}$/H$^{+}$ & $-$3.770 & \nodata & $-$4.273 & $-$3.440 \\
     O$^{+2}$/H$^{+}$ & $-$3.921 & \nodata & $-$3.353 & $-$4.200 \\
     Ne$^{+}$/H$^{+}$ & \nodata & \nodata & \nodata & \nodata \\
     Ne$^{+2}$/H$^{+}$ & $-$5.367 & \nodata & $-$3.955 & $-$5.900 \\
    \midrule
    \multicolumn{5}{c}{Physical parameters} \\
    \midrule
    $T_\text{e}$ [K]  & 8800 & 9700 & 11\,200 & 8600 \\
    $N_\text{e}$ [cm$^{-3}$] & 11\,900 & 7586 & 8500 & 2023 \\
    $c(\mathrm{H\beta})$  & 0.34 & 0.37 & 0.55 & 0.40 \\
    $\log(F(\mathrm{H\beta}))^{a}$  & $-$9.24 & $-$11.33 & $-$10.09 & $-$10.29 \\
    \midrule
    Ref. & 1 & 2 & 3 & 4 \\
    \bottomrule
    \end{tabular}
    \begin{tablenotes}
        \item $^{a}$In units of erg\,s$^{-1}$\,cm$^{-2}$ .
        \item $^{b}$From optical recombination lines.
        \item \textbf{Refs.} (1) \citet{Sharpee2004} -- (2) \citet{Shaw1989} -- (3) \citet{Ali2019} -- (4) \citet{Aleman2019}
    \end{tablenotes}
    \end{threeparttable}
    }
\end{table}

\section{Analysis and Discussion}
\label{sec:discusion}

In the following subsections, we present the analysis and discussion of the
different methods used to fit the FUV spectral region of the PNe
Hen~2-5 and Tc~1. The first method uses the parametric extinction curve of \citetalias{Fitzpatrick2007}, while the second one implements a type of HAC-like grains into a radiative transfer and photoionization code. Finally, several relevant notes regarding the PNe IC~418 and IC~2501 are provided.

\subsection{Fitting the parametric FM07 extinction law}
\label{subsec:fitting_fm07}

In order to fit the parametric extinction curve of \citetalias{Fitzpatrick2007} to the observed UV spectra, we assume that the Hen~2-5 and Tc~1 spectra are only composed by the stellar continuum (i.e. the nebular continuum does not contribute to the stellar continuum as seen in Section~\ref{sec:analysis}).
Then, we fitted the
stellar atmosphere model, assuming the $T_\mathrm{eff}$ and $\log g$ from
Table~\ref{tab:pne_iue}, to the observed IUE spectra using,
\begin{equation}
\label{eq:f_lam}
    f_\mathrm{\lambda} = \theta F_\mathrm{\lambda}10^{-0.4E(B-V)[k(\lambda - V) + R(V)]},
\end{equation}
where $\theta = (R/D)^2$ ($R$ is the stellar radius and $D$
is the distance to the CSPN), $F_\mathrm{\lambda}$ is the stellar atmosphere model (in units of erg\,s$^{-1}$\,cm$^{-2}$\,{\AA}$^{-1}$), {\ebv}
is the interstellar reddening, $R(V)$ is the ratio of reddening to extinction at $V$, and
$k(\lambda - V) \equiv E(\lambda - V)/${\ebv} is the normalised extinction curve.
The \citetalias{Fitzpatrick2007} normalised extinction curve, $k(\lambda - V)$, in the UV,
is composed by a linear component
underlying the entire UV range, a Lorentzian-like component for the
UV Bump, and a quadratic component for the FUV. This is,
\begin{equation}
\label{eq:k_lam_v}
\resizebox{0.9\columnwidth}{!}{$
    k(\lambda - V) = \begin{cases}
          c_{1} + c_{2}x + c_{3}D(x, x_{0}, \gamma), & x \leq c_{5} \\
          c_{1} + c_{2}x + c_{3}D(x, x_{0}, \gamma) + c_{4}(x-c_{5})^{2}, & x > c_{5}
    \end{cases}
    $}
\end{equation}
with
\begin{equation*}
    D(x, x_{0}, \gamma) = \frac{x^{2}}{(x^{2}-x_{0}^{2})^{2} + x^{2}\gamma^{2}}
\end{equation*}
where $c_{i}$ are the average extinction curve parameters as defined in Table~5 of \citetalias{Fitzpatrick2007}, $x$ is the wavenumber (in $\mu \mathrm{m}^{-1}$), and
$x_{0}$ and $\gamma$ are related to the position and width, respectively, of the Lorentizian-like feature representing the UV Bump.

\begin{figure}
    \centering
    \includegraphics[width=0.9\columnwidth]{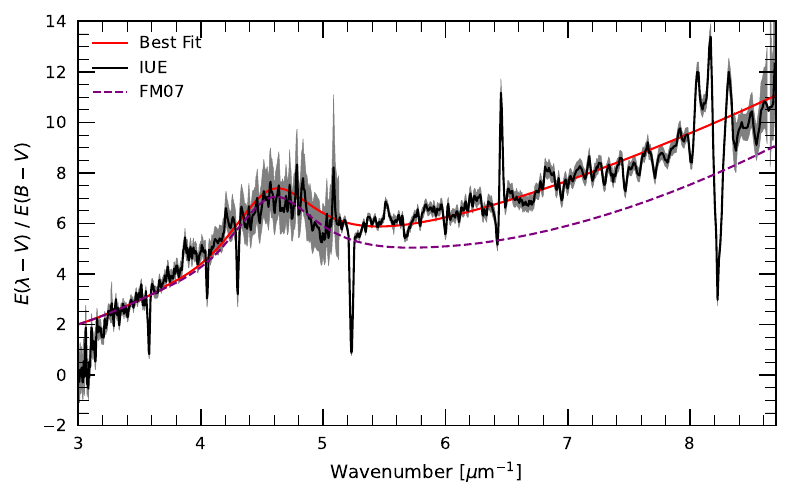}
    \includegraphics[width=0.9\columnwidth]{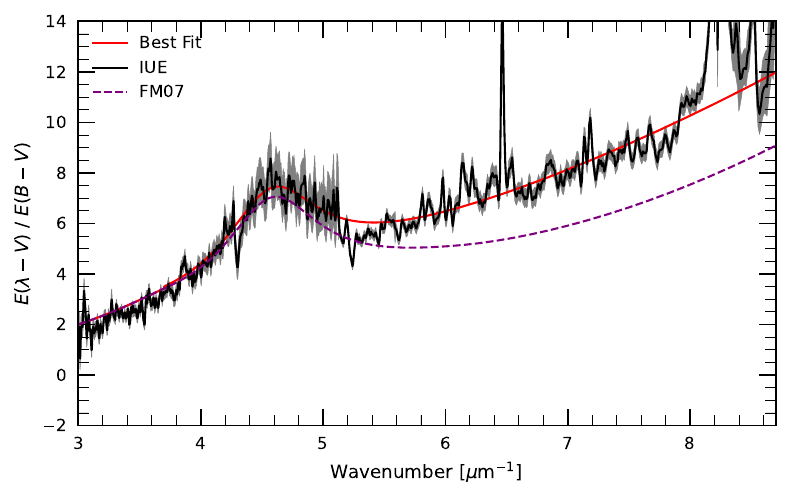}
    \caption{Extinction curves using the best fit of the parametric extinction curve of \citetalias{Fitzpatrick2007} (red line) to the Hen~2-5 (top) and Tc~1 (bottom) IUE spectra. The average extinction curve of \citetalias{Fitzpatrick2007} is shown as a reference (purple dashed-line). The extinction curve using the IUE spectra are also shown (black lines).}
    \label{fig:bestfit_fm07}
\end{figure}

Figure~\ref{fig:bestfit_fm07} shows the best fit (red line) of the \citetalias{Fitzpatrick2007} parametric extinction
curve to the IUE spectra (black line) of Hen~2-5 and Tc~1,
along with the original \citetalias{Fitzpatrick2007} (purple dashed-line) extinction curve
(using the average parameter values).
It is clear from Figure\,\ref{fig:bestfit_fm07} that the linear component ($c_{1}$ and $c_{2}$), the width ($\gamma$) and the position of the UV Bump ($c_{3}$ and $x_{0}$) are very well fitted using the average values for the extinction curve; contrary to what is observed for the quadratic component in the FUV rise ($c_{4}$ and $c_{5}$).
The best fit was then obtained by fitting the equation~\ref{eq:f_lam} to the observed IUE spectra
using the Nelder-Mead algorithm, under the
{\sc lmfit} Python package \citep{lmfit,lmfit_latest},
varying only the $c_{4}$ and $c_{5}$ parameters (which are more related to the quadratic component
responsible for the FUV rise) and {\ebv} values. 
After the Nelder-Mead algorithm, we sampled the posterior probability of the
several parameters by implementing the Markov-Chain Monte-Carlo \citep{ForemanMackey2013} algorithm,
under the {\sc emcee} Python package.
For the fitting procedure, we
reddened the synthetic spectra taking the {\ebv}, $c_{4}$, and $c_{5}$ values
as free parameters. In addition, we used priors from 0 to the maximum {\ebv}, $c_{4}$, and $c_{5}$ values as obtained for the 328 stars analysed in \citetalias{Fitzpatrick2007}, and corresponding to 1.20~mag, 0.94, and 7.50, respectively.

Table~\ref{tab:bestfit_fm07} shows that the best fit values for the parameters
$c_{4}$ and $c_{5}$ are $\sim$0.50 lower than the average values reported
in \citetalias{Fitzpatrick2007}. This corresponds to a difference of
$\sim$0.34\,mag, e.g. in the \textit{GALEX FUV} band, between the best fit and the \citetalias{Fitzpatrick2007} average extinction curve.
Such a difference could indicate that the PNe analysed here are possibly affected by a different grain size distribution towards the line of sight or that the CSPNe are extinguished by local circumstellar dust. According to the extinction analysis presented in \citet{Aleman2019}, the latter is possibly the case in Tc~1, in which, according to their study, part of the extinction is apparently internal to the nebula.

Note that the following sections are based under the hypothesis that the foreground extinction is well represented by the averaged values of the \citetalias{Fitzpatrick2007} extinction curve. However, such larger than the averaged extinction in the FUV could also be related to an accidental variation in the interstellar extinction \citep[see the sample variance in, e.g.,][]{CCM89,Fitzpatrick1999,Fitzpatrick2007} as has been proven that randomly extinction curve variations do happen as seen in the sample of \citetalias{Fitzpatrick2007}.

\begin{table}
    \centering
    \caption{Best fit extinction curve of $c_{4}$ and $c_{5}$ parameters for Hen~2-5 and Tc~1 compared to the average extinction curve parameters of \citetalias{Fitzpatrick2007}.}
    \label{tab:bestfit_fm07}
    \begin{tabular}{lccc}
    \toprule
     Name  & {\ebv} & $c_{4}$ & $c_{5}$  \\
          & (mag)  &         &          \\
    \midrule
      FM07           & 0.274 & 0.319           & 6.097           \\
      Hen~2-5   & 0.235$\pm$0.004 & 0.128$\pm$0.005 & 3.008$\pm$0.014 \\
      Tc~1        & 0.193$\pm$0.024 & 0.157$\pm$0.013 & 3.028$\pm$0.101 \\
    \bottomrule
    \end{tabular}
\end{table}

\subsection{The inclusion of HAC dust type}
\label{subsec:inclusion_HAC}

As previusly mentioned in section~\ref{sec:observations}, all sample PNe  presented in 
Table~\ref{tab:pne_iue} are classified as C-rich dust type PNe. Of these,
IC~418, IC~2501 and Tc~1 are confirmed to contain fullerenes 
\citep[see][and references therein]{Cami2010,GarciaHernandez2010,GarciaHernandez2011,Otsuka2014}.
\citet{GarciaHernandez2010,GarciaHernandez2011} showed that fullerenes are efficiently formed
in H-rich circumstellar environments and that both, fullerenes and PAH-like species, coexist and can be possibly formed
by the photochemical processing of hydrogenetaed amorphous carbon grains (HACs or a-C:Hs); a major
constituent in the C-rich circumstellar envelopes of evolved stars. In addition, it is actually believed (or mostly accepted) that the variations of the UV bump and the
FUV rise are mainly due to carbonaceous particles such as graphite, with the Drude model being the most used average extinction curve \citep{Fitzpatrick1999,Fitzpatrick2007}. However,
the possible contribution of more disordered species to the variations of the UV bump and FUV rise has been investigated also by introducing HACs,
soot particles \citep[and references therein]{Gadallah2011,Gavilan2016,Gavilan2017},
and even several carbon cluster structural families such as flakes, pretzels, branched and cages \citep{Dubosq2019,Dubosq2020}.

According to Figures~\ref{fig:pne_nebcont} and \ref{fig:bestfit_fm07}, the observed IUE PN spectra need a steeper FUV rise correction as compared to the average extinction curve of \citetalias{Fitzpatrick2007}. Although the exact origin of the contributors to the UV bump and FUV rise are still unclear,
\citet{Gavilan2016,Gavilan2017} have recently shown that soot-like nano-grains could be a possible UV bump carrier, while HAC-like grains only contribute to the FUV rise component.
Thus, in order to reproduce the observed PN UV spectra,
we extinguished the synthetic atmosphere models of Hen~2-5 and Tc~1 with the
average extinction curve of \citetalias{Fitzpatrick2007} (accounting for interstellar extinction\footnote{The average values of the parameterised \citetalias{Fitzpatrick2007} extinction curve were assumed in our analysis because there are no stars near (around $\sim$1\,{\degr}) to the line-of-sight of our C$_{60}$-PNe (e.g. with high-quality IUE spectra) to independently sample (or double-check) the foreground ISM extinction towards our PNe sample.}) and with the HAC-like grains experimental extinction curve of \citet{Gavilan2016} (accounting for circumstellar extinction, assuming $R(V)_\mathrm{HAC}=3.1$ for the HAC as in the FM07 extinction curve) as:
\begin{equation}
    f_\mathrm{\lambda} = \theta^{2} F_\mathrm{\lambda}10^{-0.4[\alpha(\lambda) + \beta(\lambda)]}
\end{equation}
with
\begin{equation}
\begin{split}
    \alpha(\lambda) & \equiv  E(B-V)[k(\lambda - V) + R(V)] \\
    \beta(\lambda) & \equiv  E(B-V)_\mathrm{HAC}[k(\lambda - V)_\mathrm{HAC} + R(V)_\mathrm{HAC}],
\end{split}
\end{equation}
where $k(\lambda - V)_\mathrm{HAC} \equiv R(V)_\mathrm{HAC}[A(\lambda)_\mathrm{HAC}/A(V)_\mathrm{HAC} - 1]$.

\begin{figure}
    \centering
    \includegraphics[width=1.\columnwidth]{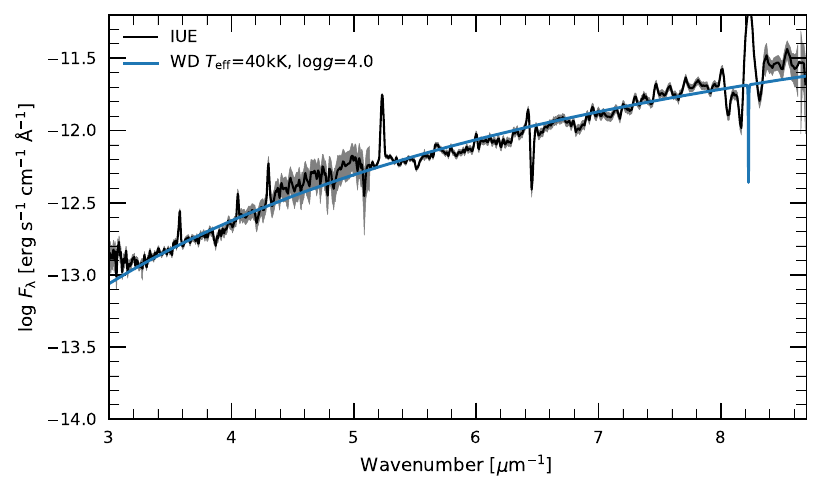}
    \includegraphics[width=1.\columnwidth]{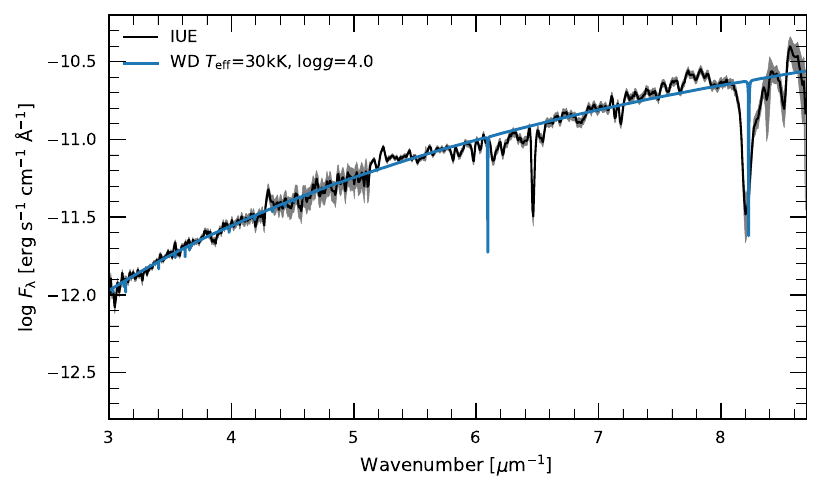}
    \caption{The IUE spectra of Hen~2-5 (top) and Tc~1 (bottom) are shown (black line) along with the best fit (blue line) using the averaged extinction curve of \citetalias{Fitzpatrick2007} in combination with the HAC extinction curve from \citet{Gavilan2016}.}
    \label{fig:bestfit_hac_cloudy}
\end{figure}

As it can be seen in Figure~\ref{fig:bestfit_hac_cloudy}, the combination of the
\citetalias{Fitzpatrick2007} and HAC extinction curves reproduce very well the observed stellar continuum of both PNe; using $E(B-V)_\mathrm{HAC}=$0.060\,mag and $E(B-V)_\mathrm{HAC}=$0.071\,mag for PN Hen~2-5 and Tc~1, respectively.

It is important to mention that the shape of the HAC extinction curve depends on whether the
HAC is more or less aromatic and on the precursor used during the experiment. 
In \citet{Gavilan2016} they produced HAC with methane (CH$_{4}$) as a precursor,
whereas in \citet{Gavilan2017} they used toluene (C$_{7}$H$_{8}$); the latter being more aromatic
than the first one. A higher aromaticity produces a weak absorption peak near 5.2{\ium} for the HAC with a toluene precursor, which is not seen for the HAC with a methane precursor neither in the IUE PN spectra presented here. 

We used the photionization code {\sc cloudy}
\citep[v22.00\footnote{\url{https://gitlab.nublado.org/cloudy/cloudy/-/wikis/NewC22}};][]{Ferland2017}
in order to analyse the effect of ionising photons and radiative transfer on the circumstellar HAC grains.
The optical constants $n$-$k$ were obtained from \citet{Gavilan2016} covering a
spectral range from 0.05{\micron} to 1{\micron}; we extended the optical constants
to lower and higher frequencies
using the BE1 amorphous carbon \citep{Rouleau1991} included in {\sc cloudy} following the
same method described in \citet{GomezLlanos2018}. A very small grains (VSG; $a_\mathrm{min}=0.001${\micron} and $a_\mathrm{max}=0.015${\micron}) distribution
was used to compile the optical constants of the HAC material, using the 'grains' command included in {\sc cloudy}, with a power law of $-$2.6 \citep{Desert1990}
resolved in a 10 bin size (generating an opacity file for the HAC grains).
We used such VSG distribution to be consistent with the work of \citet{Gavilan2016} but it is also pertinent to our C$_{60}$-PNe sample. This is because the VSGs not only are needed in order to account for the continuum emission in the near-IR for C$_{60}$-PNe like IC 418 \citep{GomezLlanos2018} but also preferentially absorb the UV light, explaining the high equilibrium temperature seen in the ISM \citep[which must be similar to that in PNe;][]{Li2001}. In addition, VSGs can also explain very well the dust temperature evolution sequence observed in C-rich PNe \citep{Stanghellini2012}. Several tests with {\sc cloudy} using different size distributions and power laws also show the need for HAC nano-grains as small as 0.001 $\mu$m (1 nm) (see Appendix~\ref{ap:gsize_dist}). For example, if we use a power law of $-$3.5 in a VSG distribution (0.001-0.015 $\mu$m), then the results are very similar, almost identical, to the results already obtained using a power law of $-$2.6; the only difference is that we would need a slightly lower amount of dust. However, by increasing the grains size distribution (i.e., 0.02-0.4 $\mu$m with a power law of $-$3.5), then the fit is not as good as the one using the VSG distribution and we would need the double amount of dust with respect to the VSG case. 

The IUE spectra were unredened with the {\ebv} values presented in Table~\ref{tab:bestfit_fm07}.
Then, we optimised the {\sc cloudy} model, including the opacity file for the HAC grains, to
reproduce the observed IUE continuum flux at 7.63, 5.71, 4.05,
and 3.30{\ium} using the \textit{phymir} method \citep{vanHoof1997}.
Abundance of other dust types such as
silicates and amorphous carbon (a-C) has been optimised as a comparison.

\begin{figure}
    \centering
    \includegraphics[width=1.\columnwidth]{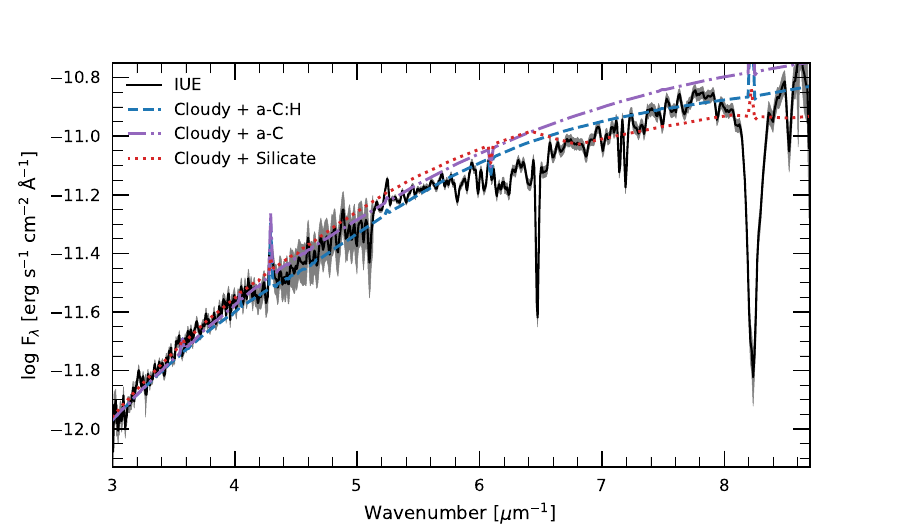}
    \caption{{\sc cloudy} models of Tc~1 optimised (see text) to the IUE spectrum (black line) with HAC (blue dashed line), amorphous carbon (a-C; purple dash-dotted line), and silicate (red dotted line). A VSG law was used in all models.}
    \label{fig:best_cloudy}
\end{figure}

We used the {\sc cloudy} model obtained in \citet{Aleman2019} for Tc~1
as an input and we only optimised the
grain abundance. Figure~\ref{fig:best_cloudy} shows the best fit obtained for
Tc~1. As it can be seen from Figure~\ref{fig:best_cloudy}, the inclusion of circumstellar HAC grains reproduce very well the stellar continuum seen in the IUE spectrum. The inclusion of circumstellar silicates fits well the FUV at $>$7{\ium} but overpredict the spectrum observed between 5.4 and 6.6 {\ium}, while the a-C circumstellar grains overpredict the spectrum observed at wavelengths longer than 5.4{\ium}. 
Note that we used {\sc cloudy} to model the observed IUE UV continuum and that we did not take into account the absorption lines see in the observed spectra (e.g., L$\alpha$ at $\sim$8.2{\ium} in Figure~\ref{fig:best_cloudy}); it should also be noted that we kept the original {\sc cloudy} continuum resolution, which means that the stellar atmosphere models were degraded to a resolution of 300.
The {\sc cloudy} program was not used to fit the Hen~2-5 spectrum because there is
insufficient chemical information available in the literature, such as
ionic and elemental abundances (or emission lines), to construct an accurate model.

As mentioned above, Tc~1 is known to contain fullerenes such as C$_{60}$ in a homogeneous ring-like structure around the CSPN \citep[see Figure 3 of][]{BernardSalas2012}. Therefore, the fact that we need a HAC component to reproduce the FUV rise absorption seen in its IUE spectrum may be related to the formation of fullerenes. In the laboratory, it is well known that the radiation-induced HAC decomposition produces both fullerenes (like the C$_{60}$ molecule) and PAH-like species as minor products \citep{Scott1996,Scott1997}; both species are indeed seen in the mid-IR spectrum of Tc~1, although Tc 1 only displays marginal PAH-like emission. From the {\sc cloudy} model fit we get that $\sim$37\% of the C in Tc 1 is in the form of HAC nano-grains. This estimate is comparable (especially considering the marginal PAH-like emission in Tc 1) to the amount of cosmic C that is estimated to be in the form of PAH-like species, $\sim$10$-$20\% \citep{Tielens2008,Li2020}, while the amount of C in the form of fullerenes like C$_{60}$ in Tc 1 is $\sim$1.5\% \citep{Cami2010}. These approximate numbers are in line with the HAC UV irradiation experiments from \citet{Scott1997} but, unfortunately, they cannot be extrapolated to the circumstellar conditions.

Although the presence of HAC-like circumstellar grains and the formation processs of fullerenes are still
unclear, \citet{GarciaHernandez2010,GarciaHernandez2011,GarciaHernandez2012} propose that the UV radiation
of the CSPN is the responsible for triggering the decomposition of circumstellar dust grains, which should have a mixed aliphatic/aromatic composition similar to that of the HAC-like dust. Based on the similarity of the IUE spectra,
this also could be the case for the PN Hen~2-5,  which shows a  very similar spectral shape in its IUE spectrum, requiring the presence of a HAC component in order to fit the FUV rise absorption. The \textit{Spitzer} mid-IR spectrum of PN Hen~2-5 is very similar to the one of C$_{60}$-PNe but it shows no clear evidence for the presence of fullerenes. At present, is not clear if the apparent non-detection of fullerenes is due to the low \textit{Spitzer} sensitivity (e.g. very weak C$_{60}$ features like the case of IC~418; see below) or to a real lack of fullerenes in Hen~2-5.

\subsection{The peculiar cases of the PNe IC~418 and IC~2501}
\label{subsec:notes_extra}

The lack of a FUV rise absorption (i.e. a HAC component) in the IUE spectra of the C$_{60}$-PNe IC~418 and IC~2501 is somehow surprising. Below we offer some notes and arguments for these peculiar cases, which could explain their different UV spectra.

The PN IC~418 is the most extended PN that displays the IR spectral signatures of fullerene and PAH-like species in the \textit{Spitzer} spectra
\citep[see][]{GarciaHernandez2012,Otsuka2014}.
\citet{DiazLuis2018} obtained N- and Q-band GTC/CanariCam mid-IR images of IC~418 in order to study the spatial distribution of fullerenes and of the carriers of some other IR dust features. They found that the fullerene emission is extremely weak and mainly concentrated in a ring-like structure around the CSPN of IC~418 with a maximum emission located towards the northeast at 6300\,au (assuming a distance to IC~418 of 1.26\,kpc).
Indeed, the \textit{Spitzer} observations just coincidentally observed the region of maximum fullerene emission\footnote{Note that the previous lower sensitivity Infrared Space Observatory (ISO) observations of IC~418 could not detect the weak C$_{60}$ features in PN IC~418 \citep{GomezLlanos2018}.}; a region that is not well covered by the IUE observations. 
However, our Figure~\ref{fig:pne_nebcont} does not show any evidence
for additional contamination by the circumstellar dust component in IC~418; i.e. there is no FUV rise absorption in the IC~418 UV spectrum and the typical ISM dust described by the average extinction curve of \citetalias{Fitzpatrick2007} can reproduce well the IC~418 UV observations.
This could be explained if the HAC circumstellar component is not homogeneous around the CSPN (and so not extinguished towards the IUE line of sight of the CSPN in the UV range), as the mid-IR GTC/Canaricam imaging seems to demonstrate \citep{DiazLuis2018}. Another explanation could be that the HAC component is already totally decomposed into fullerenes and hence we do not see any extra HAC-related contribution to the extinction of the CSPN in the UV.

Similarly to IC~418, IC~2501 is not showing any extra HAC component of extinction in the UV spectral range. However, IC~2501 is among the hottest fullerene-containing PNe with a $T_\mathrm{eff}$=55\,000\,K \citep[e.g.][]{Otsuka2014,MorenoIbanez2016}. Such a high $T_\mathrm{eff}$ would result in a high flux rate of UV photons, which could eventually destroy the circumstellar HAC-like dust or fully process the HAC grains into fullerenes like C$_{60}$ and C$_{70}$  \citep{GarciaHernandez2010,GarciaHernandez2011,GarciaHernandez2012,Otsuka2014}. 

Finally, according to the effective temperatures ($T_\mathrm{eff}$) given in Table~\ref{tab:pne_iue} and the strength of the nebular emission lines seen in our IUE spectra (Figure~\ref{fig:all_sample}), it is very likely that the youngest PN in our sample is Tc~1, followed by IC~418, Hen~2-5, and then IC~2501. Assuming that the HAC-like grains are, in part, responsible for the circumstellar extinction seen in the FUV, this could indicate that efficient HAC decomposition takes place during the early stage of the PN phase (as in Tc~1), while HACs are completely processed (converted into fullerenes and/or PAH-like species; as in IC~2501) and/or destroyed as the PN evolves toward the WD phase (i.e., for $T_\mathrm{eff}>$40 kK). However, the present statistics is very poor (only 4 PNe) and the detection of the FUV rise component in conjunction with fullerenes can also depend on how homogeneous is the circumstellar dust around the central star and how extended is the PN (as already discussed above).

\section{Summary and conclusions}
\label{sec:summary}

We have analysed the IUE spectra of IC~418, Hen~2-5, IC~2501, and Tc~1; all of them C-rich PNe and also confirmed as fullerene-containing PNe (with the exception of PN Hen~2-5).
We have obtained physical parameters and ionic abundances from
the literature in order to fit the observed UV spectra by including the nebular and the stellar
continua simultaneously in a composite model.
We found, after correcting for interstellar extinction using
the average extinction law of \citetalias{Fitzpatrick2007},
that two, Hen~2-5 and Tc~1, out of four PNe show a broad absorption feature/band at high wavenumbers
(FUV rise at $>$5.5{\ium}).
Although this extra component of the extinction seen in the PNe Tc~1 and Hen~2-5 could be possibly due to an accidental variation of the Galactic extinction curve \citep[e.g.,][]{CCM89,Fitzpatrick1999,Fitzpatrick2007}, according to \citet{Gavilan2016,Gavilan2017}, another possible explanation of such extinction component in the FUV rise could be related to the presence of HAC grains (VSG).

We fitted the UV parametric equation of \citetalias{Fitzpatrick2007} to the observed IUE spectra by varying
only the {\ebv} and the c4 and c5 parameters (correspond to the quadratic component fitted to the FUV rise).
We found that it is possible to reproduce well the observed spectra with slightly lower values of {\ebv} and with
$c_4$ and $c_5$ being $\sim$0.5 times the average parameter values from \citetalias{Fitzpatrick2007}.
This results in a \textit{GALEX FUV} magnitude difference of $\sim$0.34\,mag between using the average extinction curve and the fitted $c_4$ and $c_5$ values, which open the possibility to detect this kind of objects using the \textit{GALEX} UV colours in combination with optical-IR surveys \citep[e.g. exploring the GUVPNcat;][]{GomezMunoz2023}.
We also found that
by directly including the HAC grains extinction curve in conjunction with \citetalias{Fitzpatrick2007},
it was possible to reproduce the continuum of the IUE spectra of the PNe Tc~1 and Hen~2-5.
Similar results were obtained by using the
optical constants ($n$ and $k$ indices) ingested into the {\sc cloudy} photoionization and radiative transfer code for Tc~1. 
According to \citet{Hu2008}, HAC and soot grains, as well as carbon nanoparticles, could also explain
the unidentified IR emission (UIE) bands. However, the optical constants of HAC or soot are only available up to 1{\micron},
and we could not model the IR emission with {\sc cloudy}.
More laboratory spectroscopy efforts are strongly encouraged in order to measure the optical constants of HAC and/or soot materials in a wide wavelength range. Such measurements would play a crucial role in identifying the carriers of the UV FUV rise and the broad IR emissions simultaneously.

It is important to note that the composite model fits the observed spectra very well
using the average extinction curve for the Galaxy when considering the UV Bump and the
continuum at wavenumbers $<$5.0{\ium} in all the sample of PNe. This could suggest that there is no evidence of other carriers (different from those of the foreground ISM) such as carbon onions, affecting or contributing to the
CSM extinction in the sample PNe presented in this study.
Also, note that \citet{Aleman2019} discussed that part of the extinction found in Tc~1, by means of spatially-resolved observations, is internal to the nebula, and considering that both PNe are C-rich, this could also be the case for the PN Hen~2-5. Thus, the presence of a mixed aromatic/aliphatic hydrocarbon material, like HACs, in the CSM may be a consistent picture for C-rich PNe.
The Space Telescope Imaging Spectrograph (STIS) mounted in the Hubble Space Telescope (HST) could be used to map different regions of the PNe Tc~1 and Hen~2-5 (even close line-of-sights), by obtaining spatially-resolved long-slit spectroscopic observations (e.g., with the first order long-slit medium resolution grating). Such UV observations would provide sufficient information to precisely study the extinction variation towards both PNe, providing new additional constraints for future modelling studies.
Finally, the possible presence of both components, the HAC grains and fullerenes, in Tc~1 could support the idea that the formation of fullerenes comes from the decomposition of HACs grains through a photochemical process in the complex circumstellar envelopes of PNe.

\section*{Acknowledgements}
We acknowledge the support from the State Research Agency (AEI) of the Spanish Ministry of Science and Innovation (MCIN) under grant PID2020-115758GB-I00/AEI/10.13039/501100011033 as well as the support from the ACIISI, Gobierno de Canarias and the European Regional Development Fund (ERDF) under grant with reference PROID2020010051. This article is based upon work from European Cooperation in Science and Technology (COST) Action NanoSpace, CA21126, supported by COST.

This work is based on INES data from the IUE satellite.
The Starlink software \citep{Currie2014} is currently supported by the East Asian Observatory.

\section*{Data availability}
All data used in this manuscript are available through public archives such as INES (\url{http://sdc.cab.inta-csic.es/ines/}) and/or IUE archive (\url{https://archive.stsci.edu/missions-and-data/iue}).

\bibliographystyle{mnras} 
\bibliography{references} 

\appendix

\section{Varying R(V) to fit the FUV rise extinction seen in the C-rich PN\lowercase{e} T\lowercase{c}~1 and H\lowercase{en}~2-5}
\label{ap:rv_vary}

\begin{figure}
    \centering
    \includegraphics[width=0.95\columnwidth]{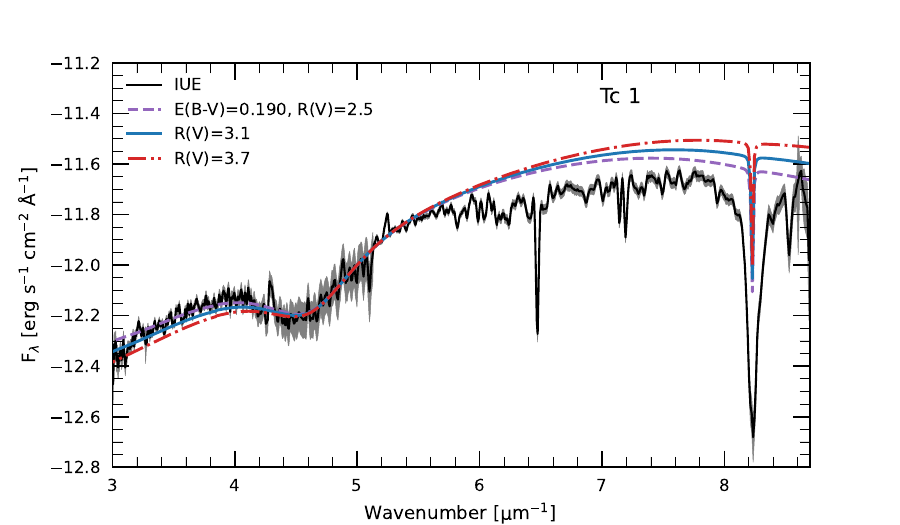}
    \includegraphics[width=0.95\columnwidth]{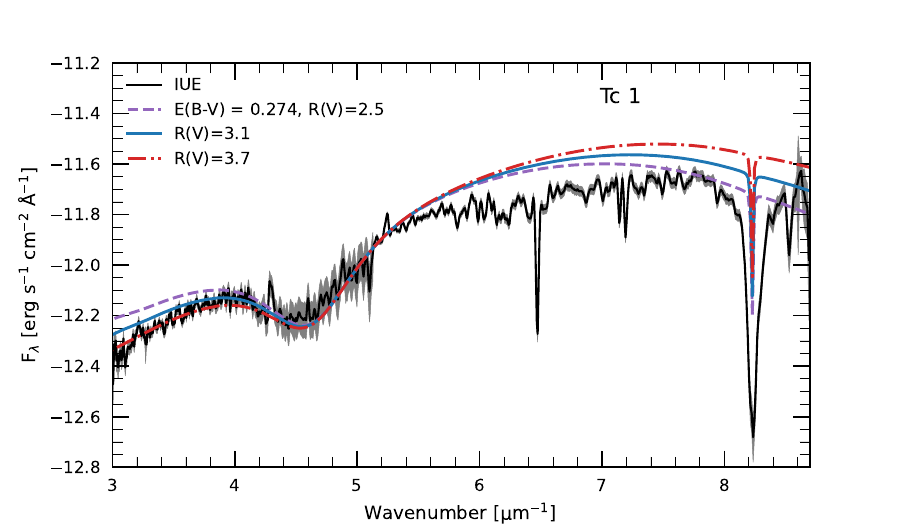}
    \includegraphics[width=0.95\columnwidth]{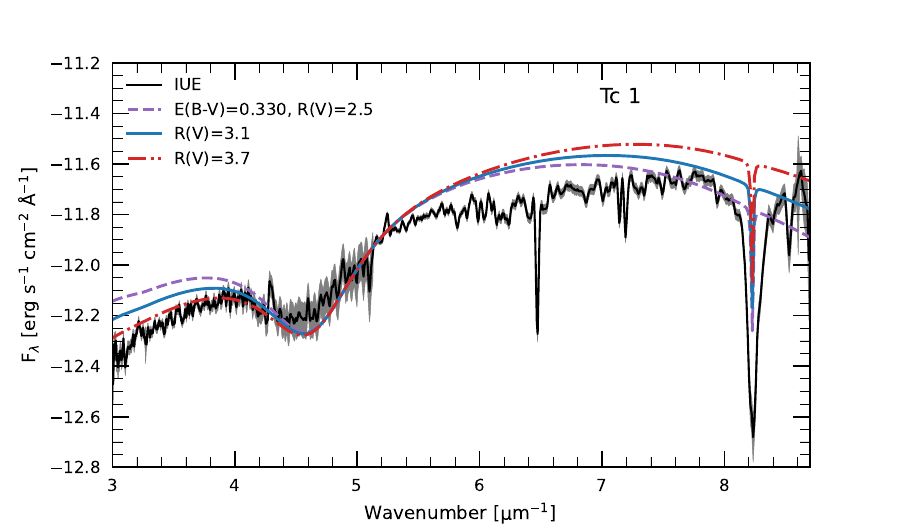}
    \caption{The IUE spectrum of Tc~1 (black line) is shown along with the stellar atmosphere model of $T_\mathrm{eff}$=30\,000\,K and $\log(g)$=4.0. The atmosphere model is reddened with $R(V)$ values of 2.5 (purple dashed line), 3.1 (blue solid line), and 3.7 (red dashed-dotted line) for three different $E(B-V)$ values of 0.190 (top-panel), 0.274 (middle-panel), and 0.330 (bottom-panel). The IUE sigma is marked as a filled grey area. \label{fig_ap:tc1_vary_rv}}
\end{figure}

\begin{figure}
    \centering
    \includegraphics[width=0.95\columnwidth]{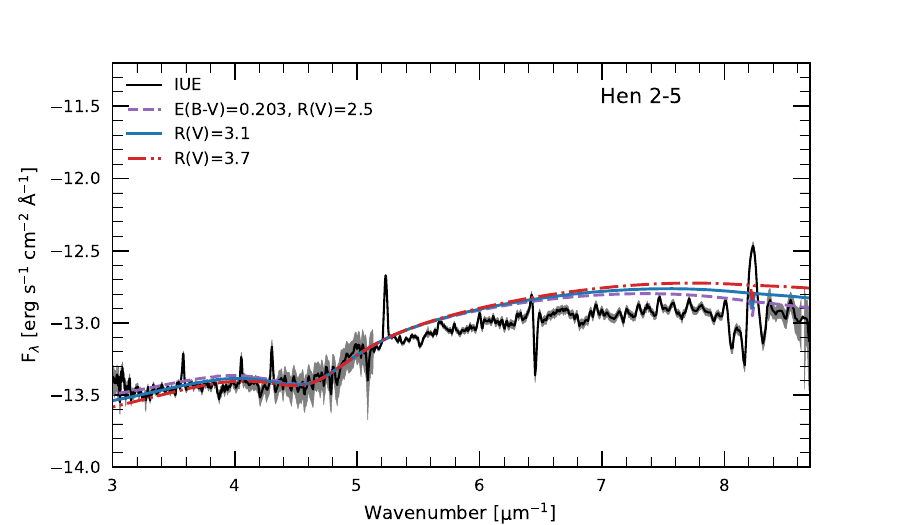}
    \includegraphics[width=0.95\columnwidth]{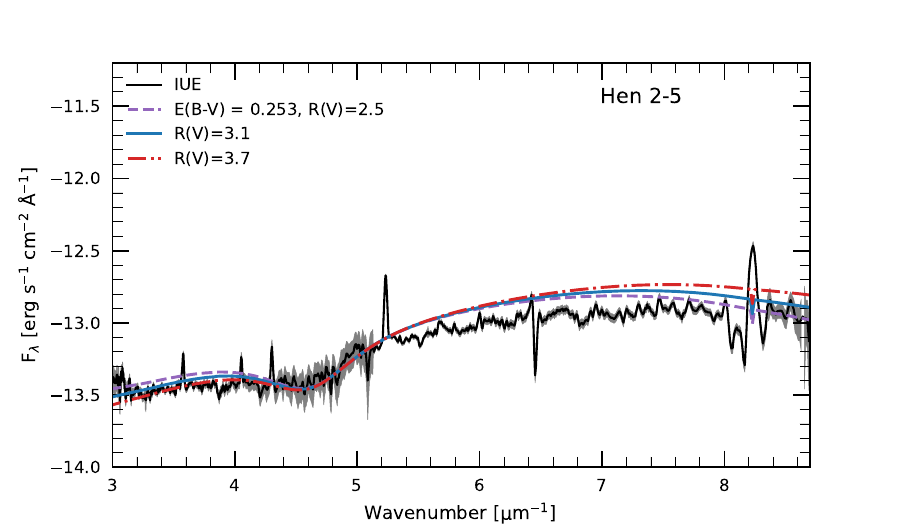}
    \includegraphics[width=0.95\columnwidth]{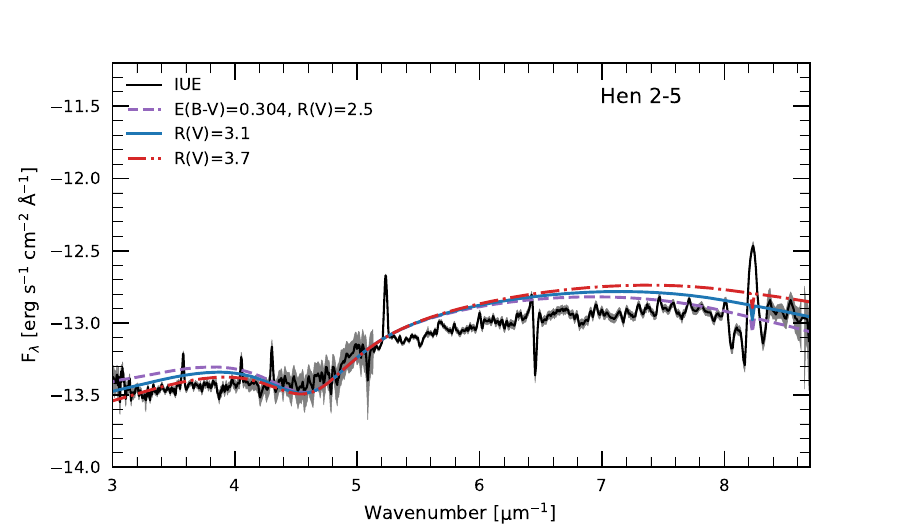}
    \caption{Same as in Figure~\ref{fig_ap:tc1_vary_rv} but for Hen~2-5, with a stellar atmosphere model of $T_\mathrm{eff}$=40\,000\,K, and with $E(B-V)$ values of 0.203 (top-panel), 0.253 (middle-panel), and 0.304 (bottom-panel).}
    \label{fig_ap:hen25_vary_rv}
\end{figure}

We use the $R(V)$-dependent extinction law of \citet{CCM89} to explore the effect on varying the ratio of total to selective extinction, $R(V)$, during the fitting procedure using the Eq.~\ref{eq:ebv_vary}. Figures~\ref{fig_ap:tc1_vary_rv} and \ref{fig_ap:hen25_vary_rv} show the minimum, maximum, and the Galactic averaged values of $R(V)$ (2.5, 3.7, and 3.1, respectively) adopted during the fitting procedure for Tc~1 and Hen~2-5, respectively. Also, a minimum, maximum, and literature value of $E(B-V)$ were considered, assuming a $\sim\pm20$\% error of the literature values (see Table~\ref{tab:physics_abundances}), as labelled in the figures. It is to be noted here that no combination of $R(V)$ and $E(B-V)$ (under the variation ranges stated above) could be found to succesfully reproduce the UV spectra observed towards both PNe Tc~1 and Hen~2-5.

\section{Effect of the HAC-like grain size distribution in the FUV rise circumstellar extinction}
\label{ap:gsize_dist}

\begin{figure}
    \centering
    \includegraphics[width=0.95\columnwidth]{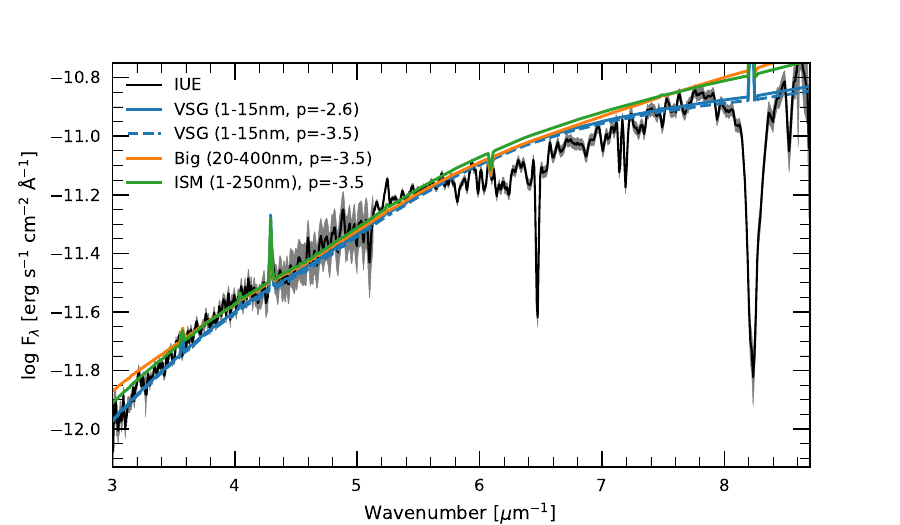}
    \caption{{\sc cloudy} models of Tc~1 optimised (see Section~\ref{subsec:inclusion_HAC}) to the IUE spectrum (black line) with HAC with the different grain size distributions labelled in the figure. \label{fig_ap:test_gr_dist}}
\end{figure}

To explore the role of the grain size distribution on the FUV rise circumstellar extinction seen in the PN Tc ~1, we run some toy models using different grain size distributions and power laws, $p$, for the HAC optical constants, $n$ and $k$ \citep{Gavilan2016}. Figure~\ref{fig_ap:test_gr_dist} shows the IUE spectrum of Tc~1 along with the different HAC grain size distributions for VSG with $p=-2.6$ (blue solid line) and $-3.5$ (blue dashed line), big grains (orange solid line) with $p=-3.5$, and canonical ISM grain distribution (green solid line; with $p=-3.5$).
Both VSG grain distributions, independently of $p$, clearly reproduce the extra extinction component seen towards the FUV in the IUE spectrum of Tc~1, contrary to what is seen when including bigger grains (as discussed in Section~\ref{subsec:inclusion_HAC}). 

The resulting C in form of dust from each HAC grain size distribution is quite different between them. The C abundance in dust form represent $\sim$37\% of the total C abundance (gas + dust phase) using the VSG ($p=-2.6$) distribution ($\sim$34\% by using the $p=-3.5$). When using only big grains, the amount of C abundance in dust form represent $\sim$80\%, the double of the value obtained when using only VSG. Finally, the combination of VSG and big grains in the ISM grain size distribution resulted in $\sim$50\% of C abundance in form of dust, which improves the fit as compared to the case of only including big grains. This test models with {\sc cloudy} using different grain size distributions show the need for HAC nano-grains as small as 1\,nm.

Finally, note that when including bigger grains, the power law exponent, $p$, becomes as important as is the parameter that shapes the grain size distribution, i.e., it weights the presence of larger versus smaller grains.

\bsp	
\label{lastpage}
\end{document}